\documentclass[aps,pre,twocolumn,showpacs,amsmath,amssymb,superscriptaddress]{revtex4-1}

\usepackage{graphicx}
\usepackage{bm}

\begin{document}

\title{Taylor Dispersion with Adsorption and Desorption}

\author{Maximilien Levesque}
\email{maximilien.levesque@gmail.com}
\affiliation{CNRS, UPMC Univ. Paris 06, ESPCI, UMR 7195 PECSA, 
75005 Paris, France}

\author{Olivier B\'enichou}
\email{olivier.benichou@upmc.fr}
\affiliation{CNRS, UPMC Univ. Paris 06, UMR 7600 LPTMC, 
75005 Paris, France}

\author{Rapha\"el Voituriez}
\affiliation{CNRS, UPMC Univ. Paris 06, UMR 7600 LPTMC, 
75005 Paris, France}

\author{Benjamin Rotenberg}
\affiliation{CNRS, UPMC Univ. Paris 06, ESPCI, UMR 7195 PECSA, 
75005 Paris, France}



\date{\today}

\begin{abstract}
We use a stochastic approach to show how Taylor dispersion 
is affected by kinetic processes of adsorption and desorption 
onto surfaces. A general theory is developed, from which we 
derive explicitly the dispersion coefficients of canonical examples like 
Poiseuille flows in planar and cylindrical geometries, both in constant 
and sinusoidal velocity fields. These results open the way for the 
measurement of adsorption and desorption rate constants
using stationary flows and molecular sorting using the stochastic resonance of the
adsorption and desorption processes with the oscillatory velocity field.
\end{abstract}

\maketitle

\section{Introduction}

In presence of a uniaxial stationary laminar fluid flow, the diffusion 
of Brownian particles along the direction of the flow is enhanced by 
an amount proportional to the inverse of the molecular diffusion 
coefficient $D_b$. This effect, known as Taylor 
dispersion~\cite{Taylor:1953,Aris:1956}, originates from the 
combination of the inhomogeneous velocity field experienced by the 
particles and the  diffusive transverse 
motion which leads to a random sampling of these fast and slow streamlines.
Taylor dispersion has implications in many fields, especially those
involving chemical reactions and determination of reaction rates such as 
microfluidics and chromatography, and  has thus been the subject of a number of works both at
the theoretical and experimental level
\cite{Van-den-Broeck:1983,C.:1990,Allaire:2010, stone_microfluidics:_2001,stone_engineering_2004,felinger:2005}.
In numerous practical situations, Taylor dispersion in the bulk flow is coupled
to the adsorption and desorption processes taking place at the walls 
confining the fluid. So far, the theoretical analysis of the resulting process 
has mainly been done explicitly in two limiting situations.

In the first class of models, the transverse motion is not explicitly 
considered, which physically corresponds to the infinitely well stirred 
limit of high diffusion coefficient $D_b$. A representative example is 
the famous  two-state model of chromatography introduced by Giddings 
and Eyring in 1955, in which a particle can be either in the mobile 
phase (in the flow) or in the immobile phase (adsorbed on the confining walls), the rates of 
change between phases being constant~\cite{Giddings:1955}. An important extension concerns 
the case when  the velocity of the mobile phase oscillates with time 
according to $v\cos(\omega t)$~\cite{Mysels:1956,C.:1982}. 
In particular, stochastic resonance  has been shown to occur if the rates 
of change between phases are both equal to $\omega/2$, leading to 
a maximum of the dispersion coefficient~\cite{Claes:1991,Jullien:2000}. 
This effect has recently proved to have applications in molecular 
sorting~\cite{Alcor2004a,Alcor2004b}.  

The second class of models has investigated explicitly the transverse motion, 
but for specific kinetics of adsorption and desorption: In~\cite{Edward:1995}, 
the dispersion coefficient is calculated when the exchanges with
the surface are infinitely fast (local chemical equilibrium), 
while Biswas and Sen have considered the situation of irreversible adsorption 
on the surface~\cite{Biswas:2007}. Besides, these studies focus on 
stationary velocity fields and the important case of oscillating velocity 
fields  mentioned above is not considered.

In this article, we develop a theoretical analysis of Taylor dispersion 
in presence of general adsorption and desorption processes. 
Relying on a stochastic approach 
(i) we derive explicit expressions of 
the dispersion coefficient  for the canonical examples of Poiseuille 
flows in planar and cylindrical geometries, both for stationary and 
oscillating velocity fields, thus opening the way to
the determination of heterogeneous rate constants from the mean velocity
and dispersion coefficient; 
(ii) we recover the fact that, in the case of a stationary velocity field, the 
 sources of dispersion associated to bulk transport 
and adsorption and desorption processes combine  additively~\cite{PRENEW_broeck_retention_1987,PRENEW_golay_1958};  
(iii) in the case of an oscillatory velocity field, we show that the 
dispersion coefficient can be optimized and discuss possible implications 
in the context of molecular sorting.


\section{The Model}

We consider a Brownian particle in a flow of velocity 
field $v$ in direction $x$.  The position of the particle in the transverse 
direction is denoted by ${\bf y}$ and the full position by 
${\bf r}\equiv(x,{\bf y})$. The longitudinal dynamics of the particle is 
assumed to be given by the Langevin equation:
\begin{equation}
\label{long1}
\dot{x}(t)=v({\bf y}(t),t)+\mathbf{1}_{b}({\bf y}(t))\eta_b(t)+\mathbf{1}_{s}({\bf y}(t))\eta_s(t),
\end{equation}
where $\mathbf{1}_{b}({\bf y}(t))$ stands for the indicator function of 
the bulk $b$ (equal to 1 if the particle's position is in the bulk 
and 0 otherwise) which accounts for bulk diffusion (with diffusion coefficient $D_b$)
and $\mathbf{1}_{s}({\bf y}(t))$  for the indicator function of the surface $s$ 
associated to surface diffusion (with diffusion coefficient $D_s$). The independent Gaussian white noises $\eta_b$ 
and $\eta_s$ are defined by their correlation functions:
\begin{equation} 
\label{long2}
 \left\{
    \begin{array}{ll}
        &\langle \eta_b(t) \rangle = \langle \eta_b(t) \rangle =0 \\
        &\langle \eta_b(t) \eta_b(t') \rangle = 2D_b\delta(t-t')\\
        &\langle \eta_s(t) \eta_s(t') \rangle = 2D_s\delta(t-t').
    \end{array}
\right.
\end{equation}
The transverse diffusion equation is driven by the evolution equation:
\begin{equation}
 \left\{
    \begin{array}{lll}
        \partial_t P({\bf y},t|{\bf y'},0) &=& D_b \nabla^2 P({\bf y},t|{\bf y'},0), \forall {\bf y}\in b \\
        \partial_t \Gamma({\bf y},t|{\bf y'},0) &=& -k_d \Gamma({\bf y},t|{\bf y'},0) + k_a P({\bf y},t|{\bf y'},0) \\
        &=&D_b \partial_n P({\bf y},t|{\bf y'},0),\forall {\bf y}\in s 
    \end{array}
\right.
\label{transverse}
\end{equation}
where $k_a$ (resp. $k_d$) is the adsorption (desorption) rate in 
length$\cdot$time$^{-1}$ (resp. time$^{-1}$) \footnote{{Note that we implicitly disregard here non-exponential waiting time
 distributions of the type studied in Ref.~\cite{compte_biased_1997}}}, $P$ (resp. $\Gamma$) is the 
propagator corresponding to a final state in the bulk (resp. on the surface)  
and $\partial_n$ stands for the normal derivative.
Initially, the particle is assumed to start from $x=0$ and the 
process ${\bf y}(t)$ to be stationary, characterized by the stationary 
distribution $P_{stat}({\bf y})$ (uniform within each phase and depending only
on the ratio $k_a/k_dL$) and the transition probability 
$P({\bf y},t|{\bf y'},t')\equiv P({\bf y},t-t'|{\bf y'},0)$.

The  first two moments of $x(t)$ are then respectively found from 
Eqs.(\ref{long1}), (\ref{long2}) to be given by
\begin{equation}
\label{eq:avgpos}
\langle x(t)\rangle = \int_0^t dt'\int_bd{\bf y} P_{stat}({\bf y})v({\bf y},t')
\end{equation}
where the integration domain of the spatial integral is the 
transverse cross section in the bulk $b$ and
\begin{eqnarray}
  \langle x^2(t)\rangle &=&
 \int_0^t dt'\int_0^tdt'' \langle v({\bf y}(t'),t')v({\bf y}(t''),t'')\rangle \nonumber\\
 &+&2D_b\langle T_b(t)\rangle + 2D_s\langle T_s(t)\rangle,
\end{eqnarray}
where $T_b(t)$ (resp. $T_s(t)$) is the cumulative time spent in the bulk 
(on the surface) up to time $t$. Finally, one has
\begin{eqnarray}
\label{dipsersion1}
&&\langle x^2(t)\rangle -\langle x(t) \rangle ^2- 2D_b\langle T_b(t)\rangle - 2D_s\langle T_s(t)\rangle=\nonumber\\
 &2&\int_0^t d\tau\int_\tau^{t}dt'\int_bd{\bf y_1} \int_bd{\bf y_2} v({\bf
y_1},t')v({\bf y_2},t'-\tau)\times\nonumber\\
 &\times&P_{stat}({\bf y_2})\left[P({\bf y_1},\tau|{\bf y_2},0)-
 P_{stat}({\bf y_1})\right] \equiv f(t).
\end{eqnarray}
We now specialize this general formula to the two experimentally
relevant cases of either a stationary or sinusoidal velocity field.
\\

\section{Case of a stationary velocity field} 

In the case of a stationary velocity field $v({\bf y},t) \equiv v({\bf y})$, 
the large time limit  of the variance of the displacement is easily 
shown from Eq. (\ref{dipsersion1}) to be given by
\begin{equation}
\langle x^2(t)\rangle -\langle x(t) \rangle ^2 \underset{t\to\infty}\sim 2K t,
\end{equation}
where the dispersion coefficient
is $K=P_{stat}(b)D_b + P_{stat}(s)D_s + K_v$, where
$P_{stat}(b) $ (resp. $P_{stat}(s)$) is the stationary probability to be in 
the bulk (resp. to be adsorbed on the surface) and
the velocity-dependent part reads:
\begin{eqnarray}
\label{generaldispersion}
K_v&=& \int_bd{\bf y_1} \int_bd{\bf y_2}v( {\bf y_1})v({\bf y_2})P_{stat}({\bf y_2})h({\bf y_1}|{\bf y_2}),
\end{eqnarray}
with 
\begin{equation}
h({\bf y_1}|{\bf y_2})\equiv \int_0^\infty \left[P({\bf y_1},t|{\bf
y_2},0)-P_{stat}({\bf y_1})\right]dt. 
\end{equation}
 Note that $h({\bf y_1}|{\bf y_2})$ is the pseudo-Green function~\cite{Barton:1989a} of the transverse problem, which  satisfies
$-D_b\nabla^2 h({\bf y_1},{\bf y_2}) = \delta({\bf y_1}-{\bf y_2}) -P_{stat}({\bf y_1})$. 
As soon as this pseudo-Green function  can be  determined, Eq.~(\ref{generaldispersion}) provides 
a general expression of the Taylor dispersion coefficient in presence of 
adsorption and desorption processes. 

Importantly, this expression can be made fully explicit in the canonical examples of planar and cylindrical Poiseuille flows, corresponding respectively to velocity fields  $v(y)=6\bar{v}\frac{y}{L}\left(1-\frac{y}{L}\right)$ (the transverse cross section being a segment of length $L$ and $y\in[0,L]$) and 
$v(r,\theta)\equiv v ( r ) =2\bar{v}\left(1-\frac{r^2}{R^2}\right)$  (the transverse cross section being a disk of radius
$R$ and ${\bf y}=(r,\theta)\in[0,R]\times[0,2\pi]$),  
where $\bar{v}$ stands for  the velocity averaged over a cross section.
The explicit determination of the pseudo-Green function $h({\bf y_1}|{\bf y_2})$ is conveniently performed by first Laplace transforming Eq.(\ref{transverse}), calculating the Laplace transform of the propagator and then going to the small Laplace variable limit. Note in particular that  the boundary conditions  Eq.(\ref{transverse})    associated to adsorption/desorption  become simple radiative boundary conditions in the Laplace domain (see for example ref.~\cite{berezhkovskii_escape_2009}). Lengthy  but straightforward calculations finally  lead to  symmetrical functions of their arguments, 
which read 
\begin{eqnarray}
\label{PFG2D}
h^\text{plan.}(y_1|y_2)&=& 
\frac{\frac{1}{2}y_1^2+\frac{k_a}{k_d}y_1 + \frac{1}{2}y_2^2 - y_2\left(L+\frac{k_a}{k_d}\right)}
     {D_b\left(L+\frac{2k_a}{k_d}\right)}
  \nonumber\\
&+& \frac{\frac{1}{3}L^3 + \frac{2D_bk_a}{k_d^2}+\frac{k_aL^2}{k_d}+ \frac{L k_a^2}{k_d^2}}
         {D_b\left(L+\frac{2k_a}{k_d}\right)^2}
\nonumber\\
\end{eqnarray}
if $y_1<y_2$ and 
\begin{eqnarray}
\label{PFG3d}
\frac{1}{2\pi}&\displaystyle\int_0^{2\pi}& d\theta_2 h^\text{cyl.}(r_1,\theta_1|r_2,\theta_2)
  = \frac{1}{2\pi D_b}\ln\frac{R}{r_2}
  \nonumber \\
  &+& \frac{\left(2R+4\frac{k_a}{k_d}\right)(r_1^2+r_2^2)-3R^3+16\frac{D_bk_a}{k_d^2}}
           {8\pi RD_b\left(R+2\frac{k_a}{k_d}\right)^2}
  \nonumber\\
\end{eqnarray}
if $r_1<r_2$.
The velocity-dependent part of the dispersion coefficients is then found to 
have the same form in both geometries:
\begin{eqnarray}
  \label{eq:Kpoiseuillebothstatic}
  K_{v}^\textrm{Pois.} &=& \alpha\frac{l^2\bar{v}^2}{D_b} 
     \frac{\beta l\left(\frac{k_a}{k_d}\right)^2 + \gamma l^2\frac{k_a}{k_d}+l^3}
          {\left(l+2\frac{k_a}{k_d}\right)^3}
+\frac{\bar{v}^2}{k_d}\frac{2l^2\frac{k_a}{k_d}}{\left(l+2\frac{k_a}{k_d}\right)^3}\nonumber\\
\end{eqnarray}
where the length $l$ and constants ($\alpha,\beta,\gamma$)
are to be substituted by $L$ and ($\frac{1}{210}$,102,18) in the planar case and 
$R$ and ($\frac{1}{48}$,44,12) in the cylindrical one.

A few comments are in order. (i) The specific case of infinitely fast exchange
with the surface (local chemical equilibrium),  considered in the cylindrical
case in  \cite{Edward:1995} as one of the generalizations of Taylor dispersion,
is  recovered  in the joint limit $k_a\to\infty,\;k_d\to\infty$ with $k_a/k_dL$ fixed, 
and is given by the first term of the right hand side (r.h.s.) of the general expression (\ref{eq:Kpoiseuillebothstatic}). (ii)  
In the infinitely well stirred limit $D_b\to\infty$, $K_{v}^\textrm{Pois.}$ is reduced to the  second  term of the 
r.h.s. of Eq.~(\ref{eq:Kpoiseuillebothstatic}).    This second source of dispersion, associated to 
adsorption and desorption kinetics only,  corresponds to the 0-dimensional dispersion coefficient used in the usual two-state model of 
chromatography~\cite{Claes:1991}, with a rate $2k_a/l$ $(l=L,R)$ for the
transition from mobile to immobile states. (iii)
Note that these two contributions to dispersion turn out to combine themselves additively.
(iv)  
The knowledge of the mean displacement $\langle x(t) \rangle \sim
P_{stat}(b) \bar{v} t$  (see Eq.~(\ref{eq:avgpos})) and the dispersion coefficient
Eq.~(\ref{eq:Kpoiseuillebothstatic}) enables the experimental determination of $k_a$ and $k_d$
from the measurement of the average velocity and dispersion coefficient for
a given species interacting with a given surface. 
(v) It also allows to design a chromatographic column for the separation of 
a mixture with known adsorption and desorption rates. 
However, in the  important case where the components of the mixture have different 
kinetic rates $k_a$ and $k_d$ but similar partitioning coefficients $k_a/k_dL$,  such a separation
is impossible with a stationary flow, since  
 $P_{stat}(b)$, hence $\langle x(t) \rangle$,  of all species are similar. As we now proceed to show, efficient sorting can nevertheless be achieved in this case by resorting to an oscillatory flow and exploiting 
the stochastic resonance between the exchange kinetics and 
the flow, thereby extending the idea put forward by Alcor \textit{et al.}
in another context~\cite{Alcor2004a,Alcor2004b}.


\section{Case of a sinusoidal velocity field}

We now consider the case where the 
velocity field is a sinusoidal function of time $v({\bf y},t)\equiv v({\bf y}) \cos(\omega t)$.
In this case, the average position $\langle x(t)\rangle$ tends to $0$ at long time,
while the function $f(t)$ in Eq.~(\ref{dipsersion1}) becomes:
\begin{eqnarray}
\label{fsinus}
&f(t)&=\int_0^t d\tau\int_\tau^{t}dt'\int_bd{\bf y_1} \int_bd{\bf y_2} 
v({\bf y_1})v({\bf y_2})P_{stat}({\bf y_2})
\times\nonumber\\
&\times&\left[P({\bf y_1},\tau|{\bf y_2},0)- P_{stat}({\bf y_1})\right]
\left[ \cos\omega(2t'-\tau) +\cos\omega\tau \right].
\nonumber\\
\end{eqnarray}
In the large time limit, the term in $\cos\omega(2t'-\tau)$ becomes negligible
and the velocity-dependent part of the dispersion coefficient is given by
\begin{eqnarray}
  \label{eq:Knonstat}
K_v=\frac{1}{2}&\displaystyle\int_b&d{\bf y_1} \int_bd{\bf y_2} 
v({\bf y_1})v({\bf y_2}) \times \nonumber \\
&&\times P_{stat}({\bf y_2}) Re(\widehat{P}({\bf y_1},-i\omega|{\bf y_2})),
\end{eqnarray}
where $Re (\widehat{P}({\bf y_1},-i\omega|{\bf y_2}))$ stands for the real part 
of the Laplace transform of the propagator, with the Laplace variable $s\equiv -i\omega$.
We focus here on the canonical  case of the planar
Poiseuille flow described above. The final result for the velocity-dependent
part of the dispersion coefficient reads:
\begin{widetext}
\begin{equation}
\label{main}
K_v^\textrm{Pois.}= 
\frac{L^2\bar{v}^2}{D_b}\frac{3 Z}{2X^6(2Y+Z)}
\frac{\chi_c^+ \cosh X + \chi_c^-\cos X + \chi_s^+\sinh X + \chi_s^-\sin X}
     {\rho_c^+ \cosh X + \rho_c^-\cos X + \rho_s^+\sinh X + \rho_s^-\sin X}
\end{equation}
where we have introduced the polynomials
\begin{equation}
 \left\{
    \begin{array}{ll}
&\chi_c^\pm=  2X^4Y^2 -6X^2Y  \pm ( X^2Z^2 +X^2 -12Y )             \\
&\chi_s^\pm=  2X^3YZ +12XY \pm ( 2X^3Y -12XY^2 -3XZ^2 -12XYZ -3X ) \\
&\rho_c^\pm=  2X^2Y^2 \pm( Z^2 +1 ) \\ 
&\rho_s^\pm=  2XYZ \pm 2XY 
    \end{array}
\right.
\end{equation}
\end{widetext}
and the reduced variables $X\equiv L\sqrt{\omega/2D_b}$,
$Y\equiv k_a/\omega L$ and $Z\equiv k_d/\omega$.
This expression constitutes one of the main results of the present work.
It extends in particular the results known in absence
of adsorption and desorption at the wall  \cite{Claes:1991} (recovered
in the limits $Y\to0$ or $Z\to\infty$) and in the case of an infinitely well stirred limit in the transverse direction (corresponding to  $X\to0$) \cite{Mysels:1956,Claes:1991,Alcor2004a,Alcor2004b}, with an adsorption rate $2k_a/L$.

Besides being an important theoretical result in itself, the expression (\ref{main}) 
also allows one to discuss the possibility of sorting components of a mixture 
in an oscillatory flow, in presence of adsorption and desorption processes.
Alcor \textit{et al.}\cite{Alcor2004a,Alcor2004b} have  demonstrated,
both theoretically and experimentally, that an oscillatory driving (by an electric field)
can be exploited to separate components of a mixture that switch between two
bulk states with different rates.
Using the 0-dimensional two-state model generally considered in
chromatography, they showed that the dispersion coefficient displays in this case a maximum when the
two rates are close to $\omega/2$, \textit{i.e.} when the average time spent in
each state are equal and comparable to a half-period of the driving, resulting 
in an effective rectification of the flow experienced by the particles. 
We now discuss the possibility of extending this idea to the case
of Taylor dispersion with adsorption and desorption.

\begin{figure}[h!]
    \includegraphics[width=0.3\textwidth]{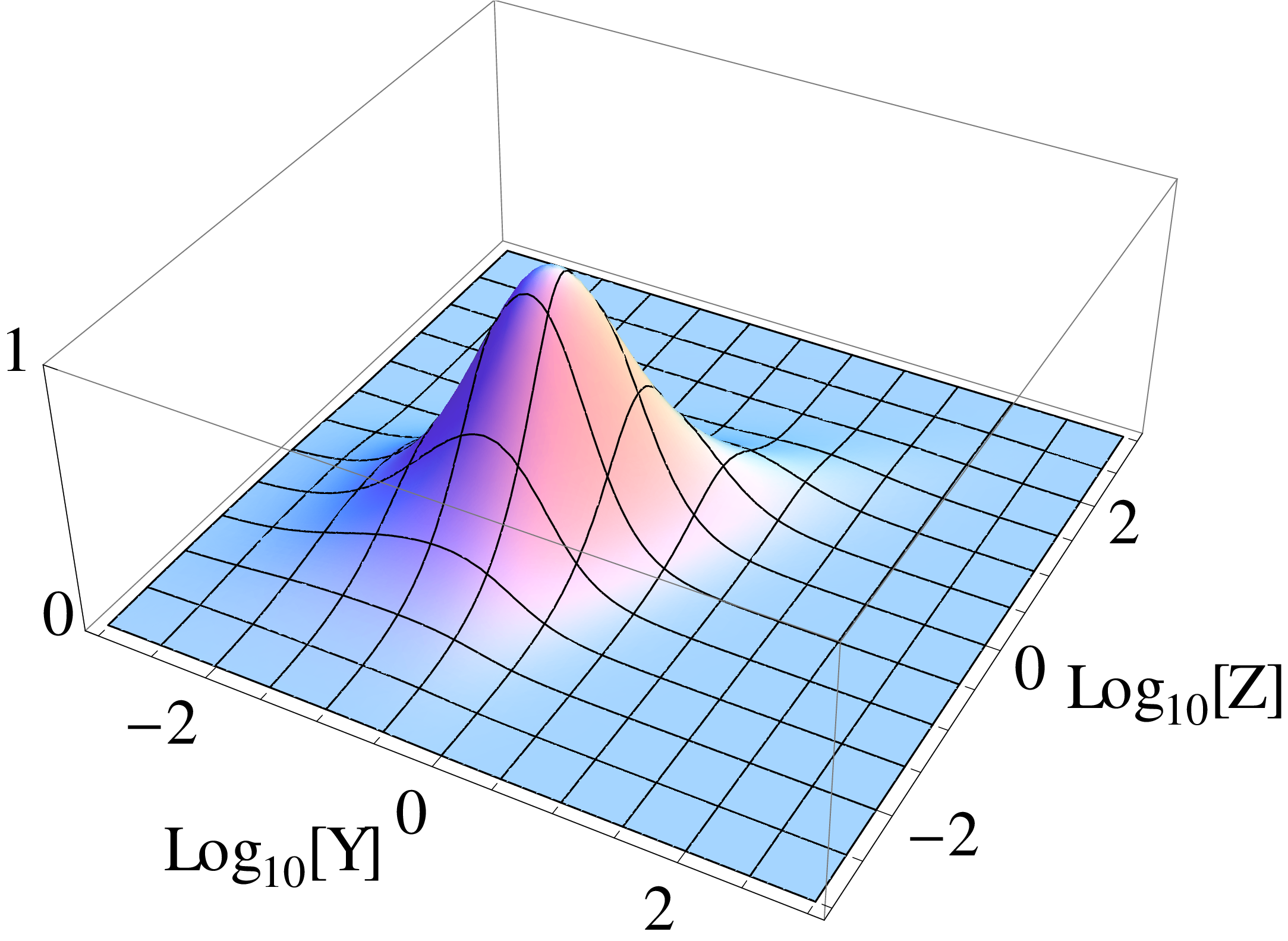}
    \includegraphics[width=0.3\textwidth]{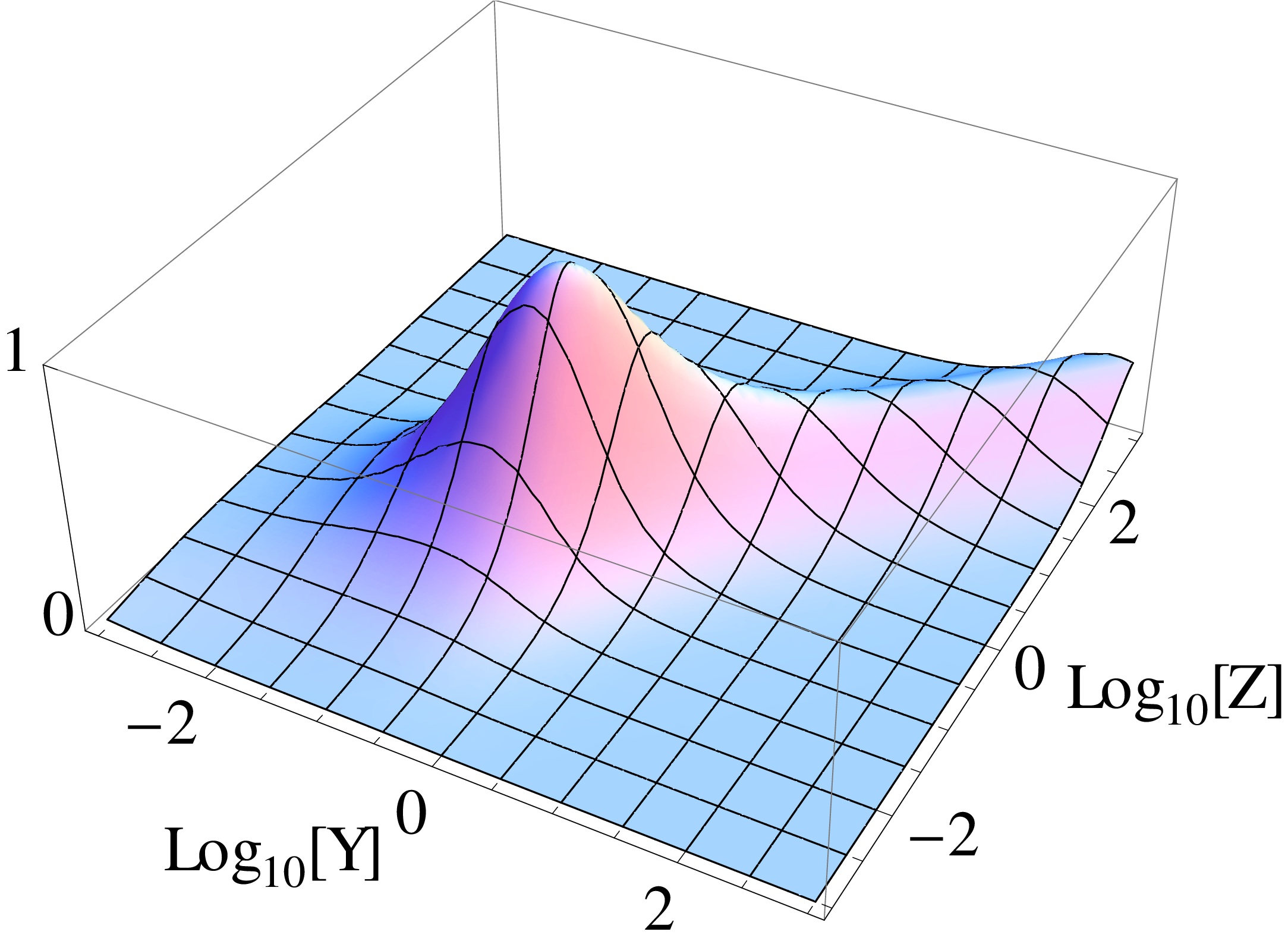}
    \includegraphics[width=0.3\textwidth]{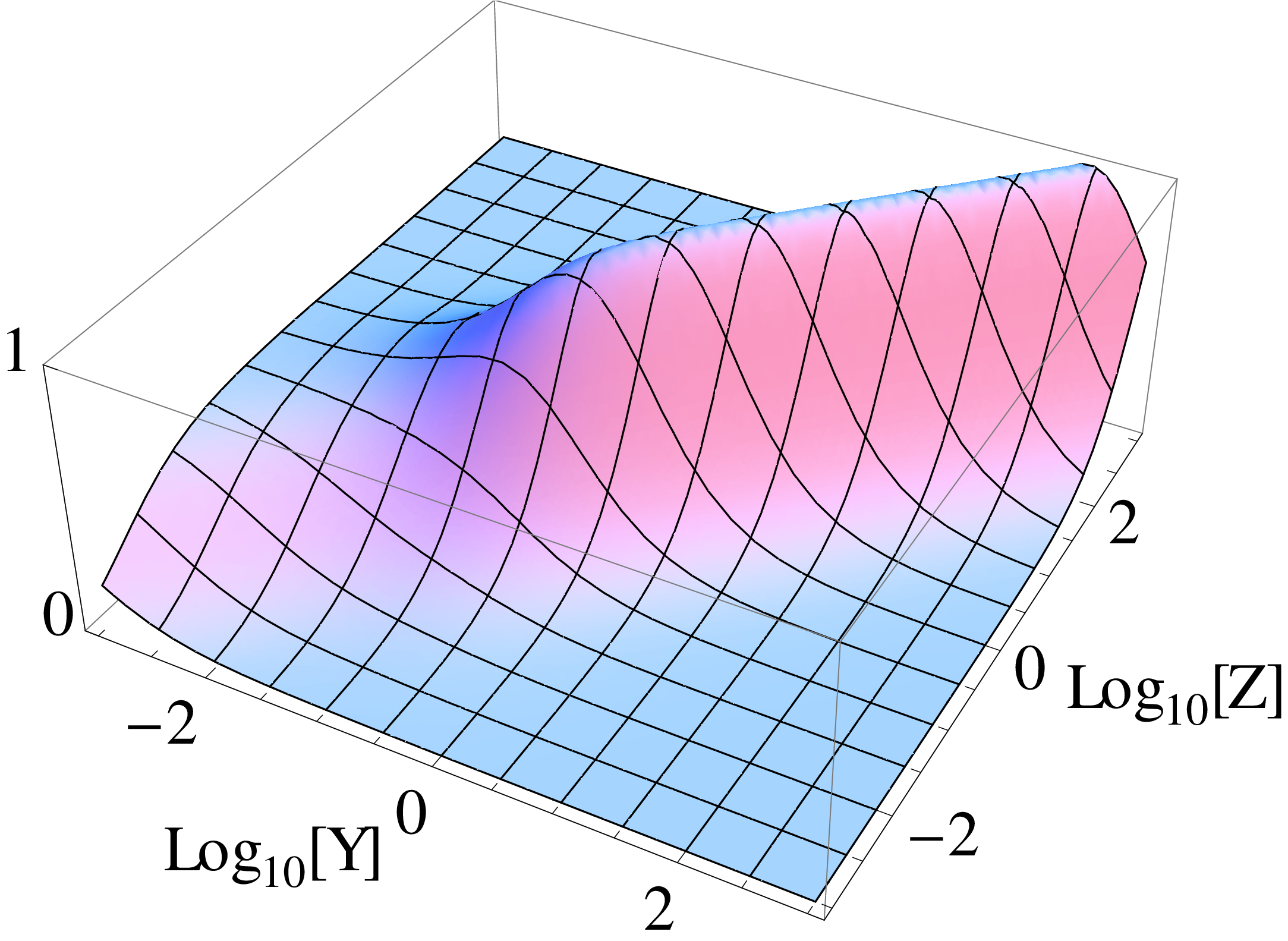}
\caption{
    \label{fig:K3D}
    (Color online) Velocity-dependent part $K_v$ of the dispersion coefficient,
normalized by the value at the optimum $K_v^{\textrm{opt}}$ as a function of
the reduced rates $Y=k_a/\omega L$ and $Z=k_d/\omega$,
for three values of $X=L\sqrt{\omega/2D_b}$: 0, 1 and 3.5.
}
\end{figure}

\begin{figure}[h!]
    \includegraphics[width=0.45\textwidth]{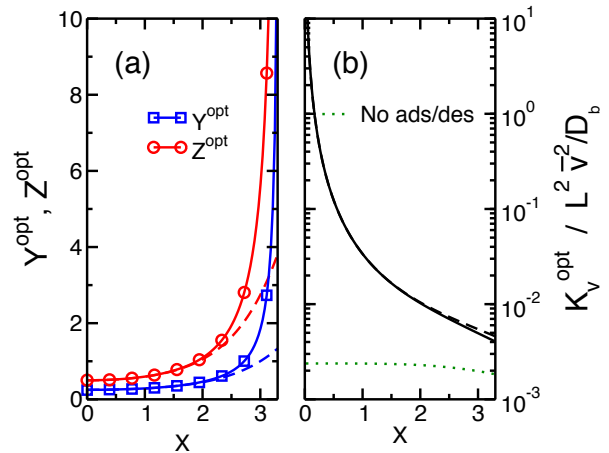}
\caption{
    \label{fig:opt}
(Color online) (a) Optimal reduced rates, $Y^\textrm{opt}=k_a^\textrm{opt}/\omega L$ 
and $Z^\textrm{opt}=k_d^\textrm{opt}/\omega$ as a function of
$X=L\sqrt{\omega/2D_b}$
maximizing the velocity-dependent part of the dispersion coefficient $K_v$.
(b) Corresponding values of the optimum $K_v^\textrm{opt}$, in units
of $L^2\bar{v}^2/D_b$.
Numerical results (solid lines) are compared to the analytical expressions 
Eqs.~(\ref{smallx_ka})-(\ref{eq:smallx_kopt}) for small $X$ (dashed lines).
In (b), $K_v$ in the absence of adsorption and desorption is also 
indicated (dotted line).
}  
\label{fig:1}
\end{figure}

Importantly, we find that Taylor dispersion with 
adsorption and desorption at the walls of the confining surfaces also exhibits a stochastic resonance.
However, the picture is more complex due to the coupling between motion 
in the direction transverse to the flow and the exchange processes.
Indeed, $K_v^\textrm{Pois.}(\omega)$
displays a global maximum for a set of finite reaction rates $(k_a,k_d)$,
provided that $X<X^*\approx3.5$, as can be seen in Figure~\ref{fig:K3D},
which reports $K_v$ normalized by the value at the optimum $K_v^\textrm{opt}$.
The optimal rates satisfy:
\begin{equation}
\label{def}
k_a^{\textrm{opt}}=\omega L Y^\textrm{opt}(X)\;\;{\rm and}\;\;k_d^{\textrm{opt}}=\omega Z^\textrm{opt}(X)
\end{equation}
where $Y^\textrm{opt}$ and $Z^\textrm{opt}$, reported in Figure~\ref{fig:opt}a, are
well approximated by their small $X$ expansions for an appreciable range (see Figure~\ref{fig:opt}):
\begin{equation}
\label{smallx_ka}
Y^\textrm{opt}=\frac{1}{4}+\frac{11}{336}X^2+\frac{163}{50400}X^4+\frac{331}{1241856}X^6+\mathcal{O}(X^8)
\end{equation}
and 
\begin{equation}
\label{smallx_kd}
Z^\textrm{opt}=\frac{1}{2}+\frac{71}{840}X^2+\frac{17}{1800}X^4+\frac{25427}{25872000}X^6+\mathcal{O}(X^8),
\end{equation}
and correspond to the following expansion of the optimal dispersion coefficient:
\begin{equation}
\label{eq:smallx_kopt}
K_v^\textrm{opt}(X)= \frac{L^2\bar{v}^2}{D_b}
 \left( \frac{1}{32 X^2} +\frac{11}{3360} - \frac{79 X^2}{564480}+\mathcal{O}(X^4)\right).
\end{equation}

While for small values of $X$, the resonance is obtained 
for $k_d\sim2k_a/L\sim\omega/2$ and has similarities
with the case of Alcor \textit{et al.}~\cite{Alcor2004a,Alcor2004b}, 
the optimal rates can in fact differ by orders of magnitude from $\omega$
as $X\to X^*$. Moreover, while the resonance process is rather selective
in the small $X$ regime, as $K_v$ decreases rapidly around the
optimum, Figure~\ref{fig:K3D} indicates that the selectivity deteriorates when
$X$ increases. Finally, the optimal dispersion coefficient, enhanced by
orders of magnitude by adsorption and desorption at the walls, also
decreases with increasing flow frequency (see Figure~\ref{fig:opt}b).
All these observations reflect the transition from the well-stirred regime
$X\to 0$ where "chemical" and Taylor dispersions combine additively
(the first two terms in Eq.~(\ref{eq:smallx_kopt}) are proportional
to $\bar{v}^2/k_d^\textrm{opt}$ and $\bar{v}^2L^2/D_b$, respectively), as in the stationary
case, to one where exchange at the walls can be limited by the diffusive influx 
of mobile species. In the latter case, only a fraction of mobile species 
can benefit from the rectification mechanism allowed by the
adsorption and desorption processes. Its effect on dispersion is then 
maximal when the particles spend the same time at the surface and in this
boundary layer accessible by diffusion within a flow period 
(now regardless of reaction rates provided that the exchange 
is fast compared to the flow frequency).

\section{Discussion}

This optimization of the dispersion coefficient can be used in the context of molecular sorting, where the problem is to isolate a species of  given $(k_a,k_d)$ from a mixture.  In fact, the ratio $Z^\text{opt}/Y^\text{opt}=k_d^\text{opt}L/k_a^\text{opt}\equiv g(X)$ 
is an increasing function of $X$, such that  $2\equiv\gamma_- < g(X)<\gamma_+\simeq 5$ for $X<X^*$,  so that 
the parameter $L$ can be chosen close to $2k_a/k_d$. This ensures that $X=g^{-1}(Lk_d/k_a)$ is arbitrarily small and determines 
the corresponding value of $\omega=2D_bX^2/L^2$ to tune.  Finally, this shows that,  in principle,  
 one can {\it always} determine parameters 
$L$ and $\omega$ such that $K_v^\text{Pois.}$ is maximal for $(k_a,k_d)$ 
while maintaining a small value of $X$ to ensure efficient sorting, 
and thus define an optimal setup to be used. Furthermore, our approach quantifies the theoretical efficiency of such sorting.

In practice, experimental constraints limit the accessible range 
of system sizes $(L_\text{min},L_\text{max})$ and flow frequencies
$(\omega_\text{min},\omega_\text{max})$. As an example, for a typical microfluidics setup,
these ranges are of the order of $(10~\mu{\rm m},1~{\rm mm})$ and 
$(0$~s$^{-1},10$~s$^{-1})$, respectively. Several cases have then to be considered:
(i)  For species such that $L_\text{min}k_d/k_a<\gamma_-<L_\text{max}k_d/k_a$, the optimal setup for efficient sorting
defined above is indeed realizable and the constraint on $\omega$ is irrelevant for all values of $D_b$; (ii) For species 
such that $\gamma_-<L_\text{max}k_d/k_a< \gamma_+$, the maximum sorting efficiency can be reached for $L=L_\text{min}$. This implies a corresponding value of $\omega=2D_bX^2/L_\text{min}^2$, which constraints the range of applicable diffusion coefficients $D_b$ to $D_b<\omega_{max} L_\text{min}^2/2X^2$. (iii) For species such that $L_\text{min}k_d/k_a>\gamma_+$ or $L_\text{max}k_d/k_a<\gamma_-$, the method is in practice not applicable.

Microfluidic technology offers a particularly versatile set of tools
to taylor the geometry, flows and surface properties to design 
experimental setups for the measurement of adsorption and desorption rates
(stationary case) or separative applications (stationary and oscillatory cases) 
according to the predictions of the present work.
In practice, for a typical microfluidic 
channel one has a transverse length $L\sim10^{-4}-10^{-3}$~m
and velocities up to $\bar{v}\sim10^{-4}-10^{-3}$~m.s$^{-1}$,
while colloids, macromolecules and molecular solutes have diffusion
coefficients in the range $D_b\sim10^{-12}-10^{-9}$~m$^{2}$.s$^{-1}$.
Oscillatory flows can be considered of the form $v({\bf y})\cos\omega t$ if 
momentum diffusion in the direction transverse to the flow is fast compared to the
period of the flow, i.e. $\omega\ll L^2/\nu$ with $\nu$ the kinematic
viscosity of the fluid. Such a condition is always satisfied 
for water in a microfluidic channel, given that 
$\nu_\textrm{H2O}\sim 10^{-6}$~m$^{2}$.s$^{-1}$ and that only frequencies smaller than
$\omega/2\pi\leq1$~s$^{-1}$ can be achieved. This last point also indicates
that the present approach will allow to measure sorption/reaction rates
slower than $1$~s$^{-1}$. As an example, the dissociation rate of DNA 
double-strands, which depends on the number of base pairs (bp), can be in
the range $10^{-5}-10^{-3}$~s$^{-1}$ for a few tens of bp~\cite{gunnarsson_single-molecule_2007,gunnarsson_kinetic_2009,volkmuth_dna_1994}.
If one considers a surface grafted with single-strand DNA, one
could selectively separate from a solution a strand containing
the complementary sequence by adjusting the flow period and the
grafting density in order to tune the adsorption rate $k_a$
(in the low surface coverage limit, it will be proportional to the latter).

\section{Conclusion}

In conclusion, the present study introduces general analytical results which
extend previous works on Taylor diffusion without adsorption and desorption
or on a 0-dimensional two-state model valid only in the perfectly stirred
limit. This approach is not limited
to the Poiseuille flows considered here as an illustration. In particular,
it can be straightforwardly extended to the case of electro-osmotic flows, 
in which an additional length scale, the Debye screening length, 
can be tuned by changing the ionic strength of the solution. The now well-established microfluidic and rising
nanofluidic technologies offer a particularly versatile set of tools
to taylor the geometry, flows and surface properties to design 
experimental setups for the measurement of adsorption and desorption rates
(stationary case) or separative applications (stationary and oscillatory cases) 
according to the predictions of the present work.

\begin{acknowledgments}
BR and ML acknowledge financial support from the Agence Nationale de la
Recherche under grant ANR-09-SYSC-012. The authors thank Magali Duvail,
Ignacio Pagonabarraga, Daan Frenkel and Pierre Levitz for discussions.
\end{acknowledgments}


\end{document}